\documentclass[aps,prl,twocolumn,showpacs,superscriptaddress]{revtex4-1}
\usepackage{amsmath,amssymb,mathrsfs}
\usepackage[dvips]{graphicx}
\usepackage{hyperref}

\def\be{\begin{equation}}
\def\ee{\end{equation}}
\def\bea{\begin{eqnarray}}
\def\eea{\end{eqnarray}}

\def\tbf{\textbf}

\def\up{\uparrow}

\begin{document}
\title{Time reversal symmetry breaking of $p$-orbital bosons in a one-dimensional optical lattice}
\author{Xiaopeng Li}
\affiliation{Department of Physics and Astronomy, University of Pittsburgh, Pittsburgh, Pennsylvania 15260, USA}

\author{Zixu Zhang}
\affiliation{Department of Physics and Astronomy, University of Pittsburgh, Pittsburgh, Pennsylvania 15260, USA}

\author{W. Vincent Liu}
\email[e-mail:]{w.vincent.liu@gmail.com}
\affiliation{Department of Physics and Astronomy, University of
Pittsburgh, Pittsburgh, Pennsylvania 15260, USA}

\date{\today}

\begin{abstract}

  We study bosons loaded in a one-dimensional optical lattice of two-fold $p$-orbital degeneracy at each site.  { Our numerical simulations find an anti-ferro-orbital p$_x$+ip$_y$, a homogeneous p$_x$ Mott insulator phase and two kinds of superfluid  phases distinguished by the orbital order (anti-ferro-orbital and   para-orbital).}  The anti-ferro-orbital order breaks time reversal symmetry.  Experimentally observable evidence is predicted for the phase transition between the two different superfluid phases.  We also discover that the quantum noise measurement is able to provide a concrete evidence of time reversal symmetry breaking in the first Mott phase.

\end{abstract}

\maketitle

The last decade has witnessed rapid experimental developments in the
studies of ultracold quantum gases in optical
lattices~\cite{2005_Bloch_lattice-review,2007_Lewenstein_OpticalLatticeReivew,2008_Bloch_LatticeReview_RMP}. Ultracold bosons and
fermions in optical lattices provide robust and controllable systems
to study correlated quantum condensed matter physics beyond the scope
of conventional many-body systems. For example, quantum phases and
quantum phase transitions of bosons in optical lattices {were studied
in recent
years~\cite{1998_Jaksch_Mott_cold-atom,2002_Greiner_mott-sf-transition,2008_Spielman_mott-sf-transition,2009_Gemelke_mott,2008_Bloch_LatticeReview_RMP,2007_Lewenstein_OpticalLatticeReivew}.

{ Besides exploring quantum phases of bosons in the lowest band, the
  possibilities of observing exotic phases of higher orbital band
  bosons {were} also put
  forward~\cite{2005_Browaeys_pband,2007_Muller_pband}.} {In more
recent experiments,} a superfluid phase with a complex order in a two
dimensional $p$-orbital band boson
system~\cite{2011_Lewenstein-Liu_orbital-dance,2011_Wirth_pband,2011_Sengstock_honeycomb_BEC}
has been {reported}, and evidences of exotic $f$-band superfluid
phases have also been
observed~\cite{2011_Olschlager_fband}. Theoretically exotic phases
predicted for both
fermions~\cite{2008_Zhao_pmott,2010_Zhang_sppair,2010_Cai_FFLO,2009_Hung_fpair,2011_Zhang_pdw}
and
bosons~\cite{2005_Isacsson_pband,2006_Liu_TSOC,2006_Kuklov_sf,2008_Lim_TSOC,2008_Vladimir_icsf,2010_Zhou_interband,2011_Li_EFA,2011_Cai_TSOC}
with orbital degrees of freedom are attracting considerable attention.
A phase-sensitive scheme of detecting the complex order of the p$_x$+ip$_y$ superfluid (SF) was proposed very recently~\cite{2011_Cai_UBEC}.

In this article, we study bosons loaded in $p_x$ and $p_y$ orbits of a one dimensional (1D) optical lattice at zero temperature with both the numerical simulations and field-theoretical methods. We find two SF phases distinguished by an orbital order---an anti-ferro-orbital (AFO) SF and a para-orbital (PO) SF, {and two Mott insulating phases---an AFO Mott and a $p_x$ Mott phase~(FIG.~\ref{fig:phasediag}).} The AFO order is a staggered orbital current ($p_x \pm ip_y$) order~\cite{2005_Isacsson_pband,2006_Liu_TSOC}. 
In the AFO SF phase, the inter-band phase difference is locked at $\pm \frac{\pi}{2}$ and the spontaneous AFO ($p_x \pm ip_y$) order in this phase breaks the time reversal symmetry (TRS), {whereas the fluctuations of the {relative phase} restore the TRS in the PO SF phase.}
Based on our results, {we propose an experimental method to distinguish different phases} by measuring momentum distribution (FIG.~\ref{fig:nk_latticedepth}), instead of directly measuring the {local current flow} resulting from TRS breaking. In this way the PO to AFO quantum phase transition, associated with TRS breaking, can be observed in experiments. The finite momentum peaks in the momentum distribution of the AFO SF phase make it distinguishable from the conventional 1D SF phases~\cite{2004_Stoferle_1DMott-SF_PRL,2004_Bloch_TG-hcb}.
In the AFO Mott phase the quantum noise measurement will be able to provide a concrete evidence of spontaneous TRS breaking.

{\it Experimental proposal.---}
{The system we shall propose is a 1D lattice elongated along the $x$ direction,} and each lattice site has a rotation symmetry in the $x$-$y$ plane (FIG.~\ref{fig:phasediag}). {In other words,} the $p_x$ and $p_y$ orbits are locally degenerate, but the hopping differs {significantly}. Such a 1D system can be realized from a 2D optical lattice. Suppose the 2D optical lattice is formed by different laser beams in the $x$ and $y$ directions and the lattice potential reads
$
V=V_x \sin^2 (k_x x) + V_y \sin^2 (k_y y)
$,
where $V_x$ {and} $k_x$ ($V_y$ and $k_y$) are the strength and wave numbers of the laser beams in the $x$ ($y$) direction. In tight binding approximation, we can use the harmonic wavefunctions to approximate the Wannier functions.
In the harmonic approximation, the local isotropy (rotation symmetry of each site in the $x$-$y$ plane) requires $V_x k_x ^2 = V_y k_y ^2$.
{This relation, which guarantees the (approximate) two-fold orbital degeneracy at each lattice site, can be somewhat surprisingly well held in the 1D limit by taking the lattice potential depth $V_y \gg V_x$ and simultaneously the lattice constants $a_y$ ($=\frac{\pi}{k_y}$) $>a_x$ ($=\frac{\pi}{k_x}$).
As a result, the local isotropy is maintained,} but the system has stronger potential and larger lattice spacing in the $y$ direction than $x$, which makes the hopping in the y direction much smaller than that in the $x$ direction. As an example, we take $V_x/E_{R,x}=6$, $V_y/E_{R,y}=24$ and $a_y/a_x=\sqrt{2}$, where $E_{R,\alpha} \equiv \hbar^2 k_{\alpha}^2/2m$ is the recoil energy in the $\alpha$ direction. Such a system is locally isotropic, while the hopping of $p_y$ orbital in the $y$ direction is smaller than one percent of the hopping of $p_x$ orbital in the $x$ direction, and the system is dynamically 1D. With the technique in Ref.~\cite{2011_Wirth_pband}, bosons can be loaded in the $p$-orbits of optical lattices.

The life time and the phase coherence of $p$-orbital bosons could be enhanced by applying double wells as in the experiment~\cite{2011_Wirth_pband}. The double well lattice gives unequal band gaps {(between the p and the lower and higher orbital bands), suppressing the decay of p-band bosons. Here, for the sake of illustrating the salient features of the 1D degenerate $p$-orbits, we employ a simple single-well lattice to simplify analysis.} Given the consistency of the experiment~\cite{2011_Wirth_pband}  with the theory on a simple square lattice~\cite{2005_Isacsson_pband,2006_Liu_TSOC}, we do not expect the complexity induced by double wells would alter the basic properties and the understanding of the reported quantum phases.

\begin{figure}[t]
\includegraphics[angle=0,width=0.6\linewidth]{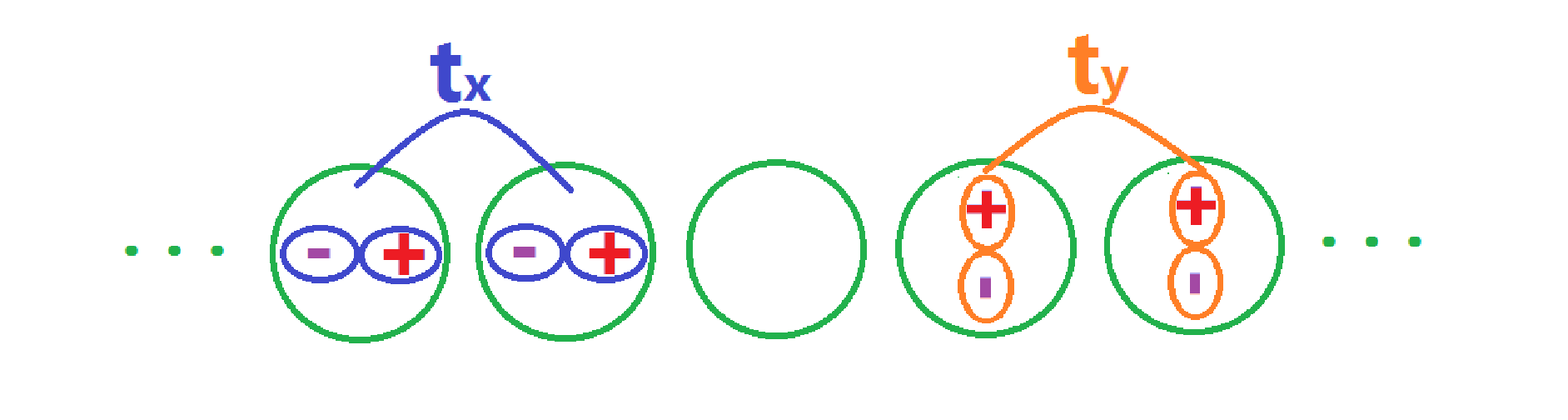}
\includegraphics[angle=0,width=0.85\linewidth]{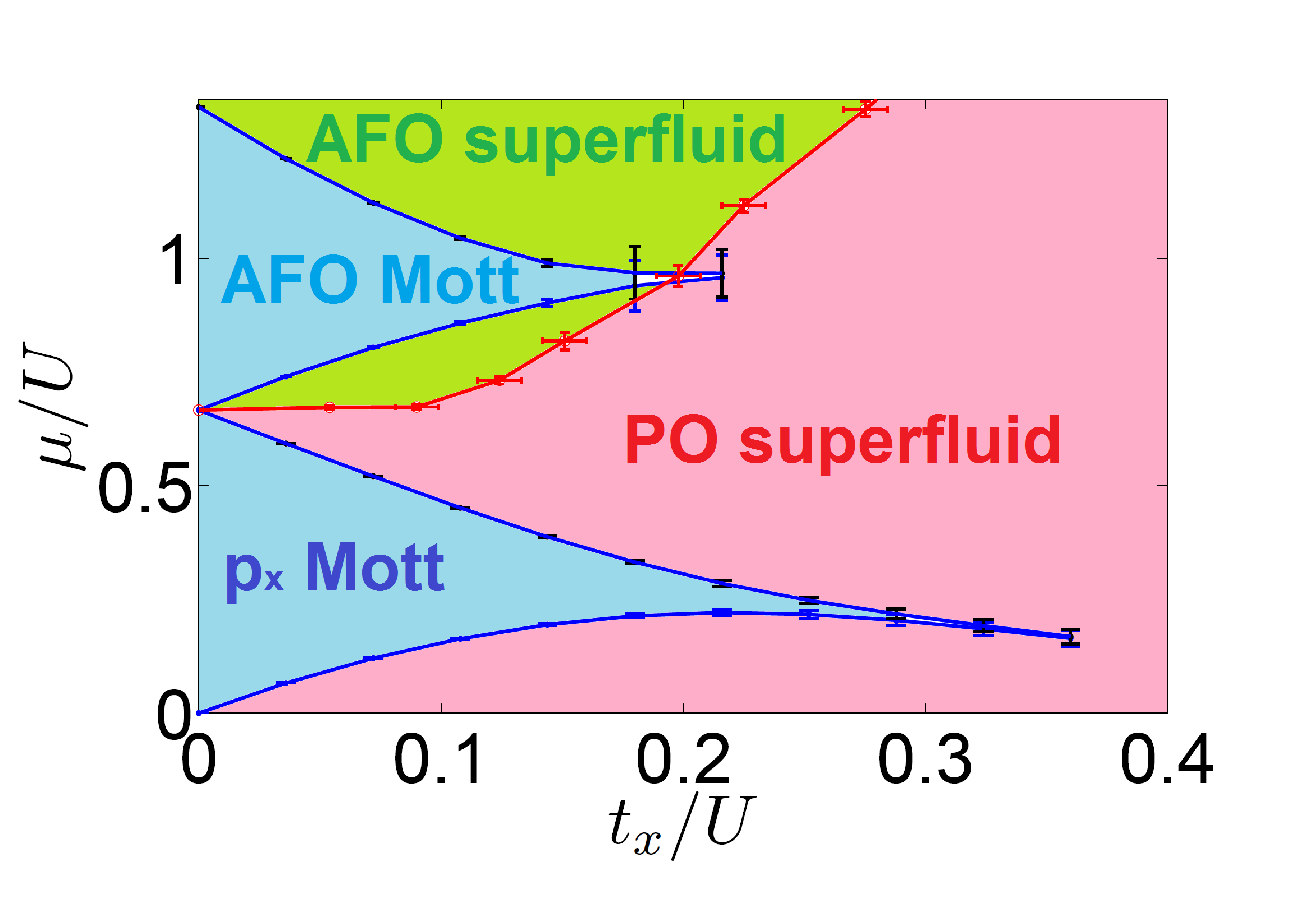}
\caption{(Color online). Phase diagram of a one dimensional lattice Bose gas with $p_x$ and $p_y$ orbital
degrees of freedom. The upper panel shows the sketch of experimental setup we proposed.
{The green circles are used to denote the requirement of the approximate local isotropy of the lattice potential at each site. }
The lowest Mott lobe (with filling $\nu=1$) is
dominated by $p_x$ bosons.
{The Mott state (with $\nu>1$) has an AFO order (see text).}
We do not claim another phase for the tiny tip of the second
Mott lobe beyond the red line because of numerical errors.
For sufficiently large hopping $t_x$ or for low filling, the Bose gas has a crossover from PO SF to a $p_x$ SF phase, which will not
be discussed in this work.
}
\label{fig:phasediag}
\end{figure}

In the 1D limit, the Hamiltonian describing bosons loaded in these $p_x$ and $p_y$ orbits reads~\cite{2006_Liu_TSOC}
\bea
\textstyle {H}  = && \textstyle \sum_{<jj'>} -t_x \hat{a}^\dag _x (j) \hat{a}_x (j') -t_y \hat{a}^\dag _y (j) \hat{a}_y (j') \nonumber \\
&& \textstyle + \sum_j \frac{U}{2} \left[ \hat{n}(j) (\hat{n}(j)-\frac{2}{3}) -\frac{1}{3} \hat{L}_z ^2 (j) -\mu \hat{n}(j)\right].
\label{eq:hamiltonian}
\eea
Here $\hat{a}_x (j)$ ($\hat{a}_y(j)$) is the annihilation operator for $p_x$ ($p_y$) orbital at site $j$.  {The discrete variable $j$ labels the sites of the 1D chain, with the lattice constant $a_x$. The local particle number operator $\hat{n}(j)$  is defined as
$
\sum_{\alpha=x,y} \hat{a}^\dag_\alpha(j) \hat{a}_\alpha (j)
$,
and the  local angular momentum operator $\hat{L}_z (j)$ is defined as
$
\sum_{\alpha,\beta} \epsilon_{\alpha \beta} (-i\hat{a}^\dag _\alpha (j) \hat{a}_\beta (j) )
$,
where the superscripts $\alpha$ and $\beta$ run over $x$ and $y$.   $U$ ($>0$) is the repulsive Hubbard interaction. The average number of bosons per site is fixed by chemical potential $\mu$. $t_x$ ($<0$) is the longitudinal hopping of $p_x$ bosons, and $t_y$ ($>0$) is the transverse hopping of $p_y$ bosons (FIG.~\ref{fig:phasediag}).
Due to anisotropy of the $p$-orbits, the longitudinal hopping (``$\sigma$ bond") is typically much larger than the transverse hopping (``$\pi$ bond")~\cite{2005_Isacsson_pband,2006_Kuklov_sf,2006_Liu_TSOC}. In this paper the ratio $|t_x/t_y|$ is taken as $9$, which corresponds to $V_x \approx 6 E_{R,x}$ (based on the estimation $ |t_x/t_y| \approx \frac{1}{2} (\pi^2 \sqrt{\frac{V_x}{E_{R,x}}} -6)$ under tight binding approximation).

\begin{figure}[t]
\includegraphics[angle=0,width=0.85\linewidth]{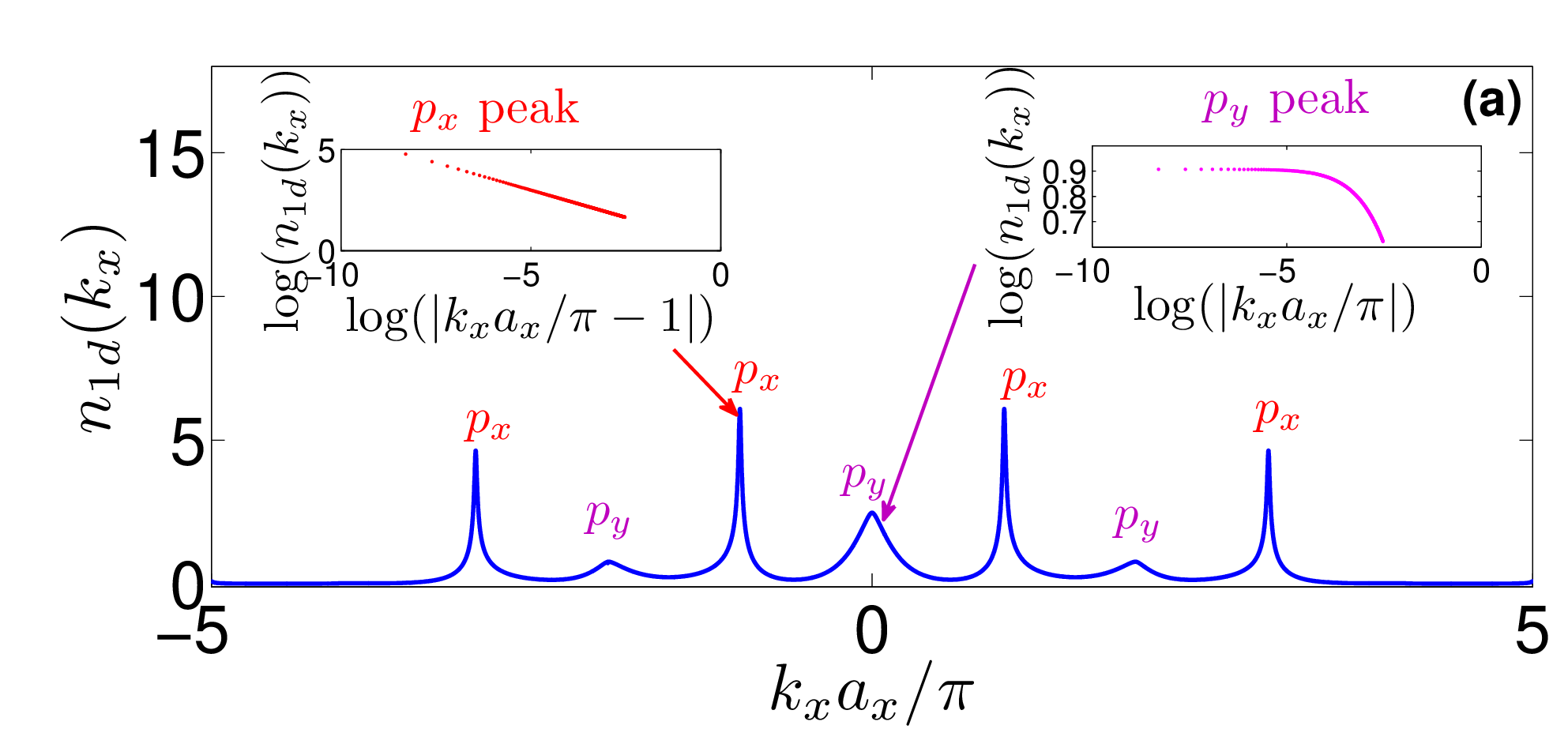}
\includegraphics[angle=0,width=0.85\linewidth]{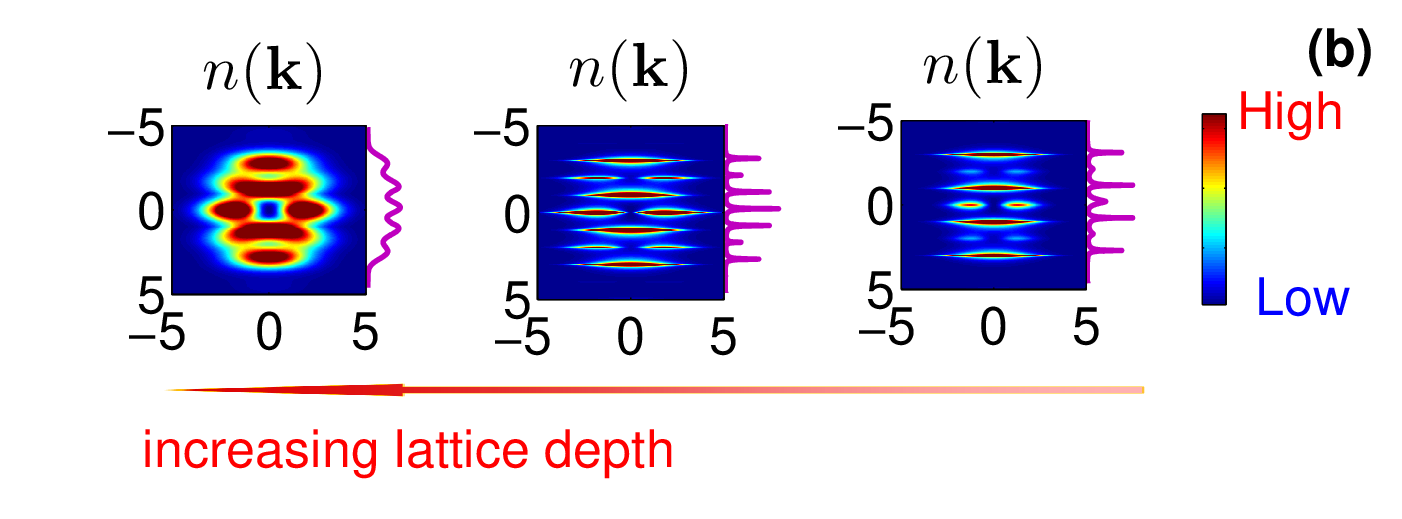}
\caption{(Color online).  Momentum distributions $\tilde{n}(\tbf{k})$. (a)
    shows the 1D momentum distribution $\tilde{n}_{1d} (k_x)$ in a geometrically 2D
    experimental setting (see paragraph ``\emph{Experimental
      proposal}'') for PO SF
    phase. $p_x$ peaks at $k_x a_x = (2j+1) \pi$ ($j$ is some integer)
    are sharp, and $p_y$ peaks at $k_x a_x = 2j\pi$ are broad. The
    insets show that the double logarithmic plot of $\tilde{n}_{1d} (k_x)$
    near the sharp (broad) peaks is linear (non-linear). (b) shows the
    sketch of 2D momentum distributions ($\tilde{n}(\tbf{k})$) in different
    phases (PO SF, AFO SF and AFO Mott from right to left). In three
    subgraphs the horizontal (vertical) axis is $k_y a_y/\pi$ ($k_x
    a_x/\pi$). The purple wiggles along each subgraph shows $\tilde{n}_{1d}
    (k_x)$. In the AFO SF phase, the $p_y$ peaks which are broad in PO
    SF, are replaced by sharp peaks.  In the AFO Mott phase, there are
    no sharp peaks.  }
\label{fig:nk_latticedepth}
\end{figure}

{The hopping term has a U(1)$\times$U(1) symmetry, which is
$\hat{a}_\alpha (j) \to [e^{i\sigma_0 \theta } e^{i\sigma_z \phi}]_{\alpha \beta} \hat{a}_\beta(j)$, with
$\sigma_0 = \left[ \begin{array} {cc}
1 & 0 \\
0 & 1
\end{array} \right]$
and
$\sigma_z = \left[ \begin{array} {cc}
1 & 0 \\
0 & -1
\end{array} \right]$. It appears that the particle numbers of $p_x$ and $p_y$ components,
$N_x =\sum_j n_x (j) $ and $N_y = \sum_j n_y (j) $ are separately conserved.
However the pair hopping term $\hat{a}_y ^\dag \hat{a}_y ^\dag \hat{a}_x \hat{a}_x$ from $\hat{L}_z ^2$
does not conserve $N_x$ and $N_y$ separately, and thus breaks the U(1)$\times$U(1) symmetry.
Only the total particle number $N = N_x + N_y$  is conserved. The U(1)$\times$U(1) symmetry is reduced to
U(1)$\times$Z$_2$ defined as
$\hat{a}_\alpha (j) \to [e^{i \sigma_0 \theta} e^{i \sigma_z \frac{\pi}{2}}]_{\alpha \beta}
\hat{a}_\beta (j)$.}

{\it Numerical Methods.---}
We use a matrix product state to represent the ground state.
The ground state is obtained by iterative optimization~\cite{2005_Schollwock_DMRG}. White's correction is implemented to avoid potential trapping in the iteration procedure~\cite{2005_White_DMRG}.
An open boundary condition is adopted in this work. The good quantum number we used in our numerical calculation is the total particle number $N$. The chemical potential is calculated as the energy it takes to add a particle (hole) to the many-body state~\cite{1998_Kuhner_onedbh,2000_Kuhner_onedbh}.
The Mott gaps are determined by extrapolating to the thermodynamic limit.
The largest system studied has $120$ sites, which is large enough to compare with the experiments on 1D quantum gases.
With the {numerical} method the ground state phase diagram of Hamiltonian in Eq.~(\ref{eq:hamiltonian})
is mapped out and shown in FIG.~\ref{fig:phasediag}.
The phase boundary of  the Mott insulating phase is determined by the vanishing of Mott gap.
The phase boundary between the AFO and PO SF phases is determined by the vanishing of
the Z$_2$ order parameter, defined as
$\tilde{L}_z \equiv \frac{1}{L} \sum_j \langle e^{iQj} \hat{L}_z (j) \rangle$, with ${Q} = \pi$.
We use a system of $40$ sites to determine this phase boundary.
The existence of the AFO and PO SF phases is verified for a system with up to $100$ sites.
The central focus of this paper is the finding of unexpected quantum orbital phases in a 1D optical lattice.
A more accurate calculation of phase boundaries is left for future study.

{\it Mott phases.---}
For the Mott phases (FIG.~\ref{fig:phasediag}), the filling factor
$\nu =\langle \hat{n}(j) \rangle$ at each site is commensurate. The
occupation number for each orbit, both $\langle \hat{a}^\dag_x  \hat{a}_x \rangle$
and $\langle \hat{a}^\dag_y \hat{a}_y \rangle$ are incommensurate. For filling
$\nu$ greater than $1$, the Mott phase features a complex order
$\langle \hat{a}_x^\dag (j) \hat{a}_y (j) \rangle \sim e^{iQ j} e^{i\zeta\frac{\pi}{2}}$ with $\zeta=\pm$ spontaneously chosen,
which breaks the U(1)$\times$Z$_2$ symmetry down to U(1). Equivalently this Mott state has a staggered angular momentum order $\langle \hat{L}_z (j) \rangle \sim e^{iQ j }$. The order parameter $\tilde{L}_z$ is finite.
Without loss of generality, we have assumed
$\tilde{L}_z$ is positive. {This Z$_2$ order also breaks the TRS, because finite $\tilde{L}_z$ means a finite local vortex-like current flow.} For filling $\nu$ equal to $1$, the Z$_2$ order does not exist for $|t_y| \ll |t_x|$. We call it $p_x$ Mott since $p_x$ boson dominates this Mott phase, i.e., $\langle \hat{a}_x ^\dag \hat{a}_x \rangle \gg \langle \hat{a}_y ^\dag \hat{a}_y \rangle$ for $|t_x| \gg |t_y|$.

{\it AFO superfluid phase.---}
By increasing the hopping the system
goes into the SF phase when the Mott gap closes. The system
has a phase transition from the AFO Mott phase to the AFO SF phase
(FIG.~\ref{fig:phasediag}). Since the Z$_2$ symmetry is broken in this SF phase,
it behaves like a single component SF phase far from the Z$_2$ critical point.
This AFO SF phase is thus characterized by an algebraic correlation
$\langle \hat{a}_\up ^\dag (j') \hat{a}_\up (j) \rangle \sim |j-j'|^{-K/2}$, where
$\hat{a}_\up ^\dag (j) = e^{iQj} \hat{a}_x ^\dag (j)  +i \xi \hat{a}_y ^\dag (j)$, with $\xi \in (0,1]$.
$\xi =1$ in the limit of $t_x / U \to 0$.
The phase correlations of original boson operators, defined as
$G_{\alpha \beta} (j, j') = \langle \hat{a}_\alpha ^\dag (j) \hat{a}_\beta (j')\rangle$,
are given by
$G_{xx} (j, j') \sim e^{iQ(j-j')}  |j-j'|^{-K/2}$,
$G_{xy} (j, j')\sim i e^{iQj} |j-j'|^{-K/2}$ and
$G_{yy} (j, j')  \sim |j-j'|^{-K/2} $
in the AFO SF phase. We emphasize that the TRS is broken in this phase, {because the off-diagonal correlation $G_{xy} (j, j')$ is complex.} 
{The key feature is that the power law decay ($|j-j'|^{-K/2}$) correlations ($G_{xx}$ $G_{xy}$ and $G_{yy}$) exhibit the same power exponent $K/2$}. In this phase, the {relative phase ($\varphi_-$) between the $p_x$ and $p_y$ SF components is locked.} The two components share the same $U(1)$ phase $\varphi_+$ at low energy. The Lagrangian describing phase fluctuations is
\be
\textstyle \mathcal{L} [\varphi_+] = \frac{1}{2\pi K}
    \textstyle \left[
        v_+ ^{-1} (\partial_\tau \varphi_+)^2 + v_+ (\partial_x \varphi_+)^2
        \right].
\ee
The Bose liquid is completely characterized by the sound velocity $v_+$ and Luttinger parameter $K$. For $K < \frac{1}{2}$, the AFO SF phase is stable against the periodic lattice potential; for $K> \frac{1}{2}$ at commensurate filling, {this phase is unstable and undergoes a localization transition towards the Mott phase~\cite{1988_Giamarchi_onedlocalization,1989_Fisher_bh}. The AFO order is preserved across the localization transition in our system.}

\begin{figure}
\includegraphics[angle=0,width=0.9\linewidth]{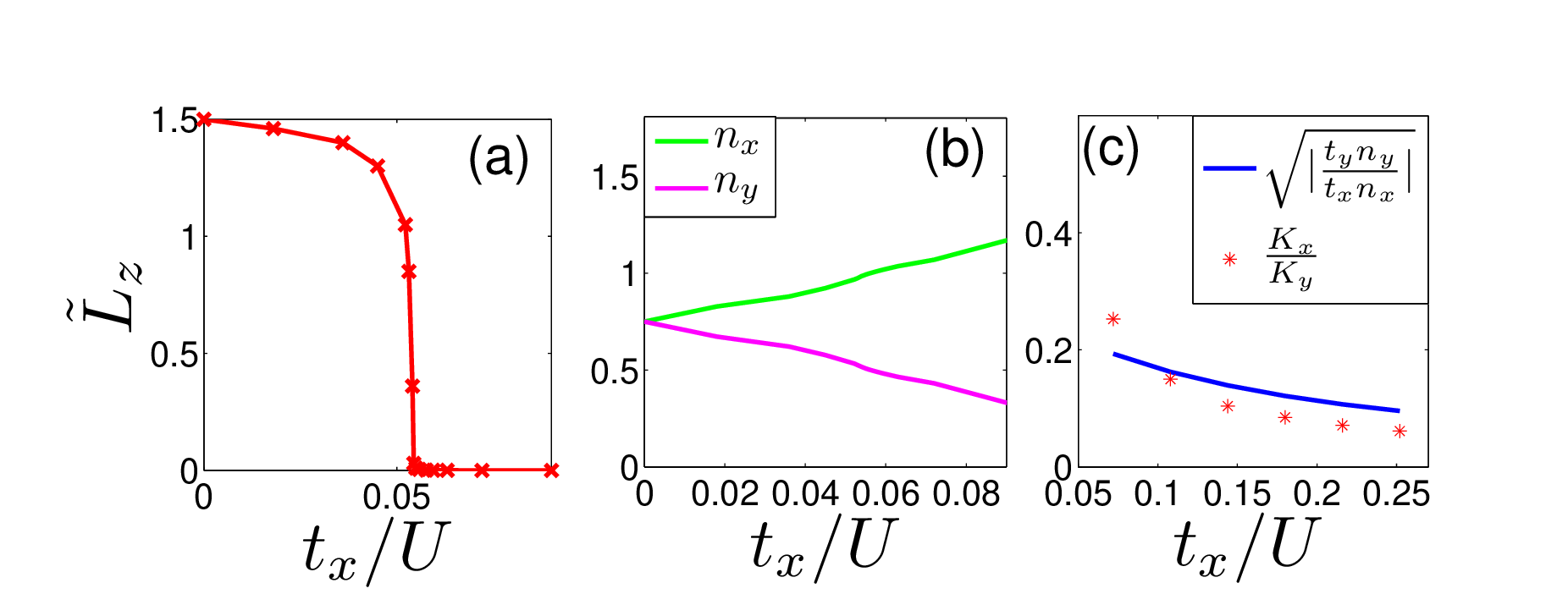}
\caption{(Color online).
Properties of the AFO to PO phase transitions with total filling $\nu=1.5$. (a) shows the Z$_2$ order parameter $\tilde{L}_z$. Our results indicate a continuous phase transition of the AFO order. (b) shows the filling of $p_x$ and $p_y$ bosons. The $p_y$ component does not vanish across the phase transition. (c) shows the ratio $K_x /K_y$ in the PO SF phase. }
\label{fig:lztransition}
\end{figure}

{\it PO superfluid phase.---}
The AFO order disappears for larger hopping, and the AFO SF gives way to the PO SF.
The behavior of the Z$_2$ order parameter $\tilde{L}_z$ and occupation numbers of $p_x$ and $p_y$ bosons across the phase transition is shown in FIG.~\ref{fig:lztransition}. {The Z$_2$ order is destroyed by quantum fluctuations of $\varphi_-$ and thus the TRS is restored in the PO SF phase.} The phase correlations of the original bosons in this phase are given by
$G_{xx}(j,j') \sim e^{iQ(j-j')} |j-j'|^{-K_x/2}$
and
$G_{yy} (j,j')\sim |j-j'|^{-K_y/2}$.
The phase coherence between $p_x$ and $p_y$ components---$G_{xy} (j,j')$---vanishes in this phase. By numerical simulations, we find $K_x \ll K_y$. Following Haldane~\cite{1981_Haldane_qliquid}, an analytic expression estimating $K_x /K_y $ is derived,
$\frac{K_x}{K_y} \approx \sqrt{|\frac{t_y n_y } { t_x n_x}|}$,
where $n_x$ ($n_y$) is the filling of $p_x$ ($p_y$) bosons. This is in qualitative agreement with the numerical results (FIG.~\ref{fig:lztransition}(c)). The Lagrangian describing this PO SF phase is
\bea
\textstyle \mathcal{L} [\varphi_x , \varphi_y]
&=& \textstyle \sum_{\alpha=x,y}  \frac{1}{2\pi K_\alpha}
    \textstyle \left[
        v_\alpha ^{-1} (\partial_\tau \varphi_\alpha)^2 +
            v_\alpha (\partial_x \varphi_\alpha)^2
    \right] \nonumber \\
    &&\textstyle + \lambda (\partial_\tau \varphi_x) (\partial_\tau \varphi_y),
\eea
where $\varphi_x$ ($\varphi_y$) is the phase of $p_x$ ($p_y$) SF component.
The mixing term ($\lambda$) is much smaller than the kinetic term ($\frac{1}{2\pi K_\alpha v_\alpha}$) in our system.

{\it Quantum phase transition from AFO to PO in the superfluid phases.---}
{The phase transition from the AFO SF to the PO SF is described by a sine-Gordon model of the {relative phase} $\varphi_-$.} The Lagrangian is
\be
\textstyle \mathcal{L} [\varphi_-] = \frac{1}{2\pi K_-}
    \left[ 
    v_- ^{-1}(\partial_\tau \varphi_-)^2 + v_- (\partial_x \varphi_-)^2
    \right]
    + m \cos (2 \varphi_-) ,
\ee
{where $m$ is estimated as $m \approx \frac{1}{3} U n_{x} n_{y}$.} When $m$ is greater than some critical $m_c (K_-)$, the sine-Gordon theory is in a gapped phase~\cite{2003_Giamarchi_oned}, and therefore $\varphi_-$ field is locked at one minimum of $m\cos(2\varphi_-)$. {Such an orbital gapped phase is the AFO SF phase.} When $m < m_c (K_-)$, {the sine-Gordon term $m\cos(2\varphi_-)$ is irrelevant in the sense of renormalization group,} and the theory is
in a gapless phase for which $\varphi_-$ is {unlocked}. This orbital gapless phase is the PO SF phase. We emphasize here the sine-Gordon term is not perturbative in our model.

{\it Experimental signatures.---}
Since the quantum phases we have found are characterized by distinct phase correlations, measuring the momentum distributions by time-of-flight (TOF) will distinguish different phases.
Assuming the interaction of atoms in TOF is negligible, the density obtained in TOF experiment ($n_{\text{tof}}$) measures the momentum distribution given as
$\tilde{n}(\tbf{k}) \propto \sum_{\alpha = x,y} \sum_n \tilde{G}_{\alpha \alpha} (k_x a_x + 2n\pi) \tilde{\phi}_\alpha ^* (\tbf{k}) \tilde{\phi}_\alpha (\tbf{k}) $
in our system (shown in FIG.~\ref{fig:nk_latticedepth}),
with $\tilde{G}_{\alpha \alpha} (k_x a_x)$ the Fourier transform of $G_{\alpha \alpha} (j, j')$ and $\tilde{\phi}_\alpha (\tbf{k})$ the momentum-space form of $p$-orbital Wannier functions. The 1D momentum distribution is defined as $\tilde{n}_{1d} (k_x) = \int dk_y \tilde{n}(\tbf{k})$.
The strong peaks of momentum distribution at finite momenta
$k_x = \pi/a_x \mod 2\pi/a_x$
will distinguish the AFO SF phases from conventional 1D SF phases~\cite{2004_Stoferle_1DMott-SF_PRL,2004_Bloch_TG-hcb}.
The Luttinger parameter can be estimated by measuring momentum distribution since it behaves as
$ \log (\tilde{n}_{1d} (k_x) )\sim~ (\frac{K}{2}-1)\log (|k_x \pm \pi/a_x|)
$ for $k_x$ near peaks $\pm \pi/a_x$. Including the effect of Wannier functions
$\tilde{\phi}_\alpha (\tbf{k})$, which are smooth slowly varying functions, does not
alter appreciably the form of $\tilde{n}_{1d} (k_x)$ near $\pm \pi /a_x$.
The effects of the harmonic trapping potential~\cite{2004_Kollath_1Dbosons_PRA} and the interactions in TOF,
which cause difficulties {of extracting the Luttinger parameter,} are left for future studies.

The AFO order in the Mott phase will have experimental signatures in the quantum noise measurement.
The quantum noise is defined as
$C(\tbf{d}) = \int d^2 \tbf{R} g(\tbf{R}, \tbf{d})$, with
$g(\tbf{R},\tbf{d}) =
\langle n_{\text{tof}} (\tbf{R} + \frac{1}{2}\tbf{d}) n_{\text{tof}} (\tbf{R} - \frac{1}{2}\tbf{d}) \rangle
- \langle n_{\text{tof}} (\tbf{R} + \frac{1}{2}\tbf{d}) \rangle \langle n_{\text{tof}} (\tbf{R} - \frac{1}{2}\tbf{d}) \rangle
$, and
$\tbf{R} = (R_x,R_y)$, $\tbf{d}= (d_x, d_y)$.
The brackets $\langle \, \rangle$ denote statistical averages of independently acquired TOF images in experiments~\cite{2005_Bloch_qnoise}.
For the Mott phases in our proposed 2D optical lattice, $g (\tbf{R}, \tbf{d})$ is given by
\bea
\textstyle g(\tbf{R},\tbf{d})
&=& L \textstyle  \left\{\sum_{\tbf{K}} \delta ^{(2)} \left( \frac{m}{\hbar t} \tbf{d} -\tbf{K} \right)
    \left (  \zeta_{xx} n_x + \zeta_{yy} n_y \right)^2 \right. \nonumber \\
  \textstyle &&+ \textstyle\sum_{\tbf{K}} \delta ^{(2)}
    \left( \frac{m}{\hbar t} \tbf{d} - \tbf{K} +\tbf{Q}_x +\tbf{Q}_y \right) \nonumber \\
   \textstyle &&\textstyle \left. \,\,\,\,\,\,\,\,\,
 \textstyle\times \left| \zeta_{xy} \mathcal{G}_{xy}  + \zeta_{yx} \mathcal{G}_{xy} ^*  \right|^2  \right\} ,
 \label{eq:qnoise}
\eea
{where $t$ is the time of flight and $L$ is the number of lattice sites.
Here, $\mathcal{G}_{xy}=G_{xy}(0,0)$,
$\zeta_{\alpha \beta} \sim (R_\alpha + \frac{1}{2} d_\alpha) (R_\beta - \frac{1}{2} d_\beta)$,
$\tbf{Q}_x = (\frac{\pi}{a_x},0)$, $\tbf{Q}_y = (0, \frac{\pi}{a_y})$ and $\tbf{K} = 2j_1 \tbf{Q}_x + 2j_2 \tbf{Q}_y$ ($j_1$ and $j_2$ are integers).
}
In Eq.~(\ref{eq:qnoise}) the smooth Gaussian part of Wannier functions $\tilde{\phi}_\alpha(\tbf{k})$ is approximated by a constant function, which is typical in quantum noise measurement~\cite{2005_Bloch_qnoise}.
The center of the trapped cold gas is taken as the origin of coordinates here.
The sharp peaks of $C(\tbf{d})$ at $\tbf{d} = \tbf{d}_0 \equiv \frac{\hbar t }{m} \left( \tbf{K} -\tbf{Q}_x -\tbf{Q}_y \right)$ signify that the off-diagonal term $\mathcal{G}_{xy}$ is finite, {which distinguishes the AFO Mott state from the $p_x$ Mott.}
The experimental signature of an imaginary $\mathcal{G}_{xy}$ is predicted to be that $g(\tbf{R}, \tbf{d})$ exhibits nodal lines at $\tbf{R} \parallel \tbf{d}_0$.
$\mathcal{G}_{xy}$ being imaginary tells a local vortex-like current flow,
which is a concrete evidence for the TRS breaking.

We are greatly indebted to I. Bloch for helping initiate this work during our overlap with him at the KITP and  for critical reading of the manuscript upon completion. We also appreciate the very helpful discussions with H.-C. Jiang, I. I. Satija and E. Zhao. This work is supported in part by {ARO (Grant No. W911NF-11-1-0230)} and  DARPA OLE Program through a grant from ARO (W911NF-07-1-0464). This research is supported in part by NSF Grant No. PHY05-51164 at KITP UCSB where this work was initiated. X.L. acknowledges the support from A. W. Mellon Fellowship.

\bibliography{onedporbital}
\bibliographystyle{apsrev}
\end{document}